# A 4-channel microfluidic hydrodynamic trap for droplet deformation and coalescence in extensional flows

*Shweta Narayan[1], Davis B. Moravec[2], Andrew J. Dallas[2] and Cari S. Dutcher[1]\**

[1]*Department of Mechanical Engineering, University of Minnesota – Twin Cities, MN 55414*

[2]*Donaldson Company Inc., Bloomington, MN 55431*

*\*Corresponding author*

**Abstract**

Two-phase liquid-liquid systems are prevalent in a range of commercial and environmental applications. Understanding the behavior of liquid-liquid systems under various processing conditions requires the study of droplet dynamics under precisely controlled flow fields. Here, we trap and control the position of droplets to study their dynamics using hydrodynamic forces alone without an external field. The hydrodynamic trap is adapted from a previously implemented 'Stokes trap' by incorporating a drop-on-demand system to generate droplets at a T-junction geometry on the same microfluidic chip. We then study confined droplet dynamics in response to perturbation by applying a millisecond-pressure pulse to deform trapped droplets. Droplet shape relaxation after cessation of the pressure pulse follows an exponential decay. The characteristic droplet shape relaxation time is obtained from the shape decay curves for aqueous glycerol droplets of varying viscosities in the dispersed phase with light and heavy mineral oils in the continuous phase. Systems were chosen to provide similar equilibrium interfacial tensions (5-10 mN/m) with wide variations of viscosity ratios. It is found that the droplet shape relaxation shows a strong dependence on droplet radius, and a weak dependence on the ratio of dispersed to continuous phase viscosity. The relaxation time is smaller for the highest viscosity ratios, potentially indicating that the dominant viscosity controls the droplet shape relaxation time in addition to the interfacial tension and droplet size. Droplet shape relaxation time can be used inform the response of droplets in an emulsion when subjected to transient flows in various processing conditions. Finally, an application of this platform for directly visualizing individual droplet coalescence in a planar extensional flow is presented. The microfluidic four-channel hydrodynamic trap can thus be applied for studying fundamental physics of droplet deformation and droplet-droplet interactions on the micro-scale to provide an enhanced understanding of emulsion behavior on an individual droplet level.

# 1. Introduction

Emulsions, or suspensions of one fluid in another immiscible fluid, are present in varied applications, including oil recovery [1,2], food processing [3–5], atmospheric aerosols [6], drug delivery [7–9] and cosmetic products [10–12]. A fundamental understanding of the physical and chemical conditions leading to emulsion stability and breakup is essential for improving processes involving emulsions. Emulsions are subjected to complex processing conditions such as pumping, mixing, extrusion or spraying leading to shear or extensional flow fields [13]. For example, micrometer-sized water droplets can become entrained in diesel fuel in automobile engines. Once entrained, the droplets can cause corrosion and pitting of components[14]. In order to remove these droplets, coalescing filters are used, where the drops are subjected to complex shear and extensional flows under confinement. During this and other emulsion processing conditions, individual droplets can be distorted, interact or coalesce with other droplets. Better understanding of the response of a confined droplet to a perturbation in shape can yield insights into the emulsions' dynamics in confined flows. In this work, we identify three key areas whereas studies of droplet response during emulsion processing are under-characterized, namely, 1) droplet response to sudden cessation of flow (characterized by a shape relaxation time) 2) effect of dispersed phase viscosity on droplet shape relaxation in liquid-liquid systems and 3) effect of confinement on droplet relaxation time.

In prior single droplet deformation studies, experimental and theoretical approaches have been used to characterize the behavior of discrete droplets suspended in another fluid phase. From a theoretical perspective, seminal papers by Lord Rayleigh [15], Sir Horace Lamb [16] and G.I. Taylor [17,18] laid the foundation for single droplet deformation studies, providing the analytical solutions for droplet behavior in emulsions. Of particular relevance to the current study are the solutions derived by Lamb [16], who obtained a theoretical expression for the damping time of a liquid globe in a fluid of negligible density and viscosity and for a spherical bubble oscillating about an equilibrium shape, both in the inviscid limit. Taking this to the next step, Chandrasekhar [19] extended this derivation for a droplet of finite viscosity in an inviscid outer phase (e.g. air) or a bubble in a fluid of finite viscosity to obtain expressions for the oscillation frequency and damping time, based on the fluid viscosity and surface tension. Most notably, Chandrasekhar identified a critical value of drop/ outer fluid viscosity above which the droplet or bubble would not oscillate but return to its equilibrium shape in an aperiodic manner exhibiting a slow decay. Suryanarayana and Bayazitoglu [20] solved Chandrasekhar's equations numerically to illustrate aperiodic (or periodic) motion in the limit of high (or low) droplet viscosity. Furthermore, Miller and Scriven [21] derived analytical expressions for oscillation frequency and damping time of a droplet of finite viscosity suspended in another quiescent medium of a finite viscosity. Finally, these cases were defined for free interfaces, inextensible interfaces or interfaces populated by insoluble and soluble surfactants by Miller and Scriven, Lu and Apfel and others [21–25]. These early works have provided the foundation for more advanced theoretical studies of droplet deformation in shear and extensional flows by Barthes-Biesèl and Acrivos [26], Cox [27] and others [28–31].

Experimentally, devices which trap droplets at a stagnation point can be applied towards measurement of the droplet's response to perturbation. Historically, 'droplet tweezers' or 'droplet traps' have been considered a highly tunable and versatile method for studying single droplet dynamics. Most techniques for capturing or trapping particles or droplets employ an external field such as optical, electric or acoustic fields to exert a trapping force. For droplet-in-air systems, optical tweezers have been employed widely for trapping droplets and particles and measuring their properties, including viscosity and surface tension, from droplet shape oscillations [32–34]. Similarly, acoustic tweezers have been employed for trapping droplet arrays and measuring material properties from drop deformation dynamics [22–24]. For liquid-liquid systems, the four-roll mill developed originally by Taylor [18] has been miniaturized and automated by Leal and coworkers [35–37] to study drop deformation under controlled flow fields. The four-roll mill has been employed extensively to study droplet deformation under steady shear and extensional flows; for example, Stone and Leal [37] employed this apparatus to investigate the transient response of droplets in a suspending medium (with varying viscosity ratios) upon sudden cessation of flow when subjected to large deformation. While the relaxation time was found to be controlled solely by interfacial tension, the qualitative features of the deformation exhibited a dependence on viscosity ratio. For instance, when drops were subjected to an initially large deformation, drop breakup was observed upon cessation of flow in low viscosity ratio cases, whereas for the highest viscosity ratio, Stone and Leal observed a slow relaxation of the droplet back to a steady spherical shape. This finding indicates that the viscosity ratio dictates droplet deformation dynamics when drops are subjected to flow transients.

Droplet microfluidic devices have also been used for a wide range of experimental studies on droplet dynamics including deformation, coalescence and breakup. Christopher et al. used a simple T-junction microfluidic geometry to study coalescence and breakup of slug-shaped droplets as a function of Capillary number, finding a critical Capillary number for coalescence and splitting which scales with slug curvature and viscosity ratio [38]. Similarly, flow through a microfluidic cross-slot device has been employed to study slender shapes of highly viscous droplets as a function of Capillary number [39]. Deformation of droplets flowing through a microfluidic contraction or expansion has also been extensively used as a method to measure interfacial tension between two fluid phases [6,14,40–42]. The effect of solid particles at the liquid-liquid interface on droplet coalescence has been studied by colliding droplets generated at opposing T-junctions by Zhou et al. [43]. Coalescence and breakup of a large number of droplets in an emulsion has also been studied using geometric expansions, cross-slots or contractions in microfluidic devices [44–46]. While these microfluidic methods provide valuable information about droplet deformation and coalescence, they lack precise control over droplet position and are conducted in strong bulk flow environments, such that the velocity of fluid in the outer phase strongly influences droplet dynamics.

A microfluidic version of the four-roll mill was developed by Hudson et al [47], capable of generating both extensional and shear flows. Schroeder and coworkers first developed a microfluidic hydrodynamic trap, which employed a two-layer microfluidic device (with a control layer and a fluidic layer) with feedback control for trapping particles at a stagnation point in a cross-slot using on-chip membrane valves [48–50]. The microfluidic hydrodynamic trap was improved further by Shenoy et al [51] by replacing the two-layer membrane valve device with a single-layer microfluidic device, where trapping is achieved by controlling the flow rates in the channels forming the cross-slot using an optimization algorithm for control. This device, called a 'Stokes trap' can be used for gentle hydrodynamic trapping and steering of particles in the microfluidic device. Moreover, the number of stagnation points in a Stokes trap can be increased by increasing the number of channels, such that two stagnation points can be employed to trap two particles simultaneously to study particle collisions [51,52]. The Stokes trap has previously been implemented for studying vesicle and polymer dynamics at a stagnation point [53,54]. Ramachandran and coworkers employed a diamond-shaped hydrodynamic cross-slot to measure dynamic interfacial tension for bitumen-water systems as well as to study soft particle assembly in flow [52,55,56]. However, the Stokes trap to date has not been applied to studies of droplet dynamics and response to flow transients.

Finally, while theory and experiments pertaining to droplet response in an unconfined flow field abound in literature, relatively fewer studies consider the effect of confinement on a droplet's shape response to perturbation. Particularly in microfluidic devices, droplets are inevitably subjected to forces due to the confining walls of the channel, resulting in a combination of shear and extensional fields acting on droplets [40]. Previous studies on the effect of degree of confinement on the deformation of drops in a microfluidic extensional flow device have revealed that increasing confinement tends to enhance droplet deformation at a given strain rate [57]. In shear flows, confinement has been shown to slow down droplet relaxation by several researchers [58–61]. In fact, Minale et al. found that for a non-Newtonian continuous phase e.g. Boger fluid, a single relaxation time can be used for droplet shape relaxation dynamics when the drops are highly confined, unlike for unconfined drops, where two relaxation timescales have been identified [61]. Ulloa et al. [62] used a flow-through microfluidic cross-slot device to study relaxation of confined droplets in an extensional flow field and found that under confinement, it is not the viscous stress, but the pressure in the outer fluid that dominates the droplet's response to deformation. The relaxation of highly confined droplets in Hele-Shaw cells has been studied extensively using boundary element simulations and a depth-averaged model for Stokes flow by Gallaire and coworkers [63–65]. The shape relaxation dynamics of droplets in confined flow fields can be extremely critical for understanding the flow start-up and cessation transients occurring during the flow of emulsions in various processing applications, including filtration and mixing [66].

In this work, an investigation of droplet shape relaxation dynamics of droplets trapped at a microfluidic stagnation point is presented. The Stokes trap developed by Shenoy et al. [51] is modified to include on-chip LabVIEW-controlled droplet formation using a droplet-on-demand technique. Droplets formed in the device are gently trapped at the center of a cross-slot with hydrodynamic forces alone using feedback control. Next, the shape of a trapped droplet is perturbed by applying a millisecond-pressure pulse to deform the droplet while keeping the droplet trapped. Upon cessation of the pressure pulse, the droplet shape relaxation is analyzed to obtain droplet shape relaxation time, over a range of viscosity ratios and droplet sizes. With this measurement, we aim to address the key areas identified previously where current experimental investigations are lacking - 1) droplet relaxation time in relatively stagnant flow conditions 2) effect of dispersed phase viscosity on droplet relaxation time and 3) confinement effects. The measured relaxation times at various viscosity ratios and droplet sizes provide valuable information about the dynamics of droplets in micro-confined environments. Finally, an application of the hydrodynamic trap for studying binary droplet coalescence is highlighted.

## 2. Methods and Materials

*2.1 Device Fabrication and Experimental Setup*

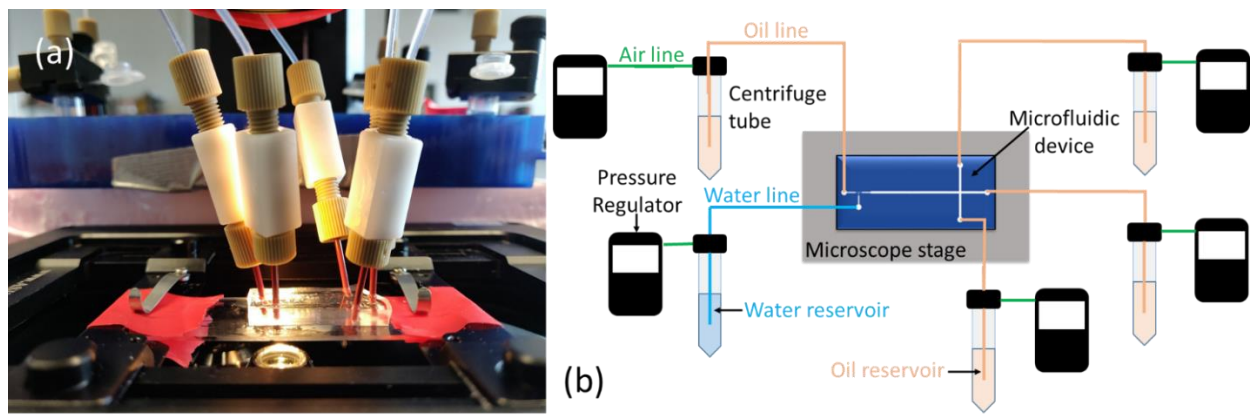

**Figure 1:** (a) Experimental setup of the hydrodynamic trap for droplets, showing the microfluidic device mounted on an inverted microscope with the rigid tubing (red) used to control flow resistance. (b) Schematic of the trapping setup, showing: 1. Microfluidic chip, 2. Microscope stage, 3. Centrifuge tubes used as reservoirs for the fluids 4. Continuous phase fluid (oil), 5. Dispersed phase fluid (aqueous), 6. Tubing used to deliver oil to the device, 7. Tubing used to deliver aqueous phase to the device, 8. Analog controlled pressure regulators and 9. Tubing used to deliver pressurized air to the reservoirs.

The microfluidic device used for hydrodynamic trapping is fabricated using standard soft-lithography methods [67–69]. Briefly, the mask design for a four-inch silicon wafer is created using DraftSight (Dassault Systèmes) and printed on transparencies (CAD/Art Services) with a resolution of 8 μm. Silicon wafers are cleaned using piranha etching ($H_2SO_4$ and $H_2O_2$ in a 3:1 ratio) at 120°C for 15 minutes, dried and spin-coated with SU-8 2050 (Microchem) to achieve a photoresist layer of height ~120 μm. After pre-baking, the photoresist is exposed to UV with the mask aligned over it using a Karl Suss MA6 Mask Aligner. The wafer is developed after a post-exposure bake using propylene glycol monomethyl ether acetate and rinsed with isopropyl alcohol. The channel depths are quantified using a profilometer (KLA-Tencor P-16). The profilometry results are included in the Supplementary Information (SI).

The microfluidic devices are made of polydimethylsiloxane (PDMS, Sylgard-184) from Dow Corning with a 10:1 ratio of base to curing agent. The silicon wafer is treated with trichlorosilane (Gelest

Inc.) in a vacuum desiccator for 20 minutes before pouring PDMS and baked at 70°C for at least 4 hours. The devices are then cut out and entry ports are punched using 1.5 mm biopsy punches (Miltex). The PDMS devices are sealed to glass slides (Thermo Fisher Scientific) by plasma treatment (Harrick Plasma) for 1 minute and baked for at least 2 hours before use. After baking, the device is rendered hydrophobic by treatment with Novec 1720 electronic grade coating (3M), which is injected into the channels. After complete evaporation of the liquid, the device is heated at 130°C for 1 hour. Furthermore, oil is injected into the device and allowed to sit overnight prior to use, to completely avoid adhesion of aqueous droplets to the device walls during the experiment.

The sealed device is then mounted on the stage of an inverted microscope (Olympus IX 83) with a calibrated stage. Two cameras are used for imaging – a Basler ace acA1300-60gm camera with a maximum frame rate of 60 fps at 1.3 MP resolution is used for trapping the droplets using LabVIEW (National Instruments), while a Photron Mini UX100 high speed camera is used to capture the rapid droplet deformation dynamics. The hydrodynamic trapping setup has previously been described in detail by Shenoy et al [51]. Pressure regulators from Proportion Air (QPV series) are used to pressurize the headspace of air in microfluidic reservoirs (Darwin Microfluidics, XS size). The requisite pressure is converted to voltage using the following conversion: The pressure range for the regulators is 0-30 psi (gauge), and the output voltage is 0-10 Volts DC. The pressure in psi (gauge) is multiplied by 10/30 to convert it to voltage and this voltage signal is sent to a National Instruments Data Acquisition board with a cDAQ-9174 chassis, NI 9264 Analog Output module and NI 9201 Analog Input module, which is in turn wired to the pressure regulators. The reservoirs hold the fluids to be supplied to the microfluidic device, and are connected to the device via Teflon tubing (1/16" OD x .020" ID, IDEX Health and Science) coupled with a small section of high-pressure drop PEEKsil tubing (IDEX Health and Science, 100 μm ID) of length 5 cm. with a union assembly (IDEX Health and Science) as shown in **Figure 1(a)**. The high-pressure drop tubing section is sized such that the resistance of this section is at least 5-10 times that of the microfluidic channels.

*2.1 Droplet-on-demand and Hydrodynamic Trapping*

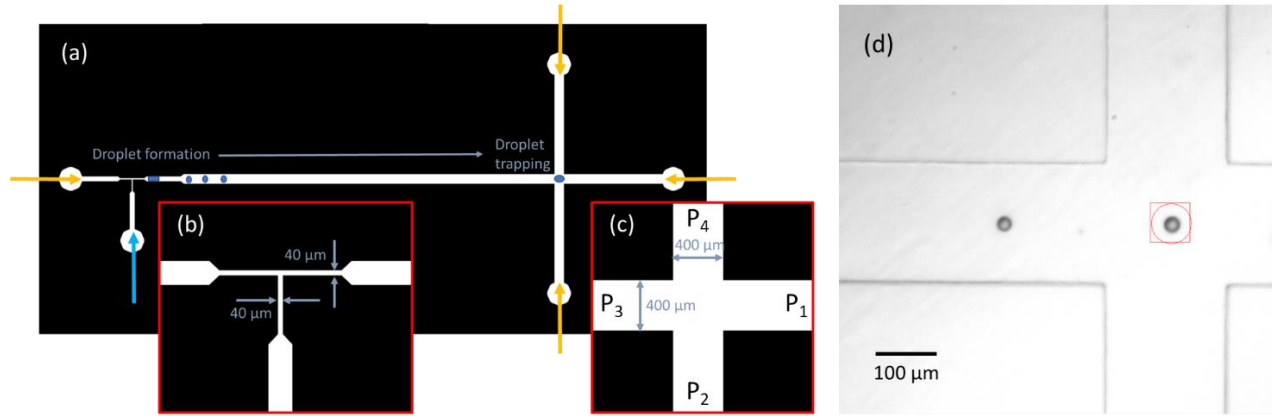

**Figure 2:** (a) Microfluidic device design for the modified hydrodynamic trap, showing the location of the T-junction for droplet formation on the microfluidic chip and the four-channel cross-slot for droplet trapping on the same chip. (b) Dimensions of the T-junction for droplet formation (c) Dimensions of the cross-slot for droplet trapping, including the designations $P_1 - P_4$ for the four channels forming the cross-slot. (d) Image of a droplet trapped in the modified hydrodynamic trap. The red square is an overlay used to indicate the desired position or 'set point' for trapping, and the red circle is used to indicate the current or actual position of the droplet.

In this work, the Stokes trap developed by Shenoy et al. [51] has been modified to allow for droplet generation on the same microfluidic chip. Essentially, a T-junction geometry is added upstream of the cross-slot to allow for generation of monodisperse droplets with controllable size and speed [70]. Several microfluidic geometries are available for generating monodisperse droplets, including T-junctions, co-flow and flow-focusing devices [70–72]. However, it was found that the T-junction geometry was best suited for the hydrodynamic trap, since it causes minimal disruption of the flow-field in the trapping region. For hydrodynamic trapping with uniform droplet generation, the pressure of the dispersed phase is comparable to the pressure in the continuous phase, which shears off the dispersed phase at the T-junction. The Laplace pressure drop across the droplet interface for a T-junction of width 40 μm and height 100 μm is on the order of 0.1 psi (gauge) for all the oil-aqueous systems studied here, meaning that maintaining the meniscus at the T-junction intersection requires the dispersed phase pressure to be approximately 0.1 psi (gauge) higher than the continuous phase. Maintaining a fluid-fluid meniscus at the T-junction intersection is critical for generating a small number of droplets (2-3) using a drop-on-demand technique. The T-junction does not play a role in trapping droplets and is placed far upstream of the cross-slot geometry to avoid interference of the droplet formation process with the trapping mechanism.

Aqueous phase droplets are generated on-demand to ensure that three or fewer droplets are generated, which can then be trapped downstream of the T-junction in the trap region. This controlled drop generation is important because the incoming droplets can often coalesce with or displace a trapped droplet in the cross-slot. If the goal of the microfluidic experiment is to trap a single droplet to study its relaxation dynamics, it is preferable to avoid interference from incoming droplets. The drop-on-demand technique is implemented using an excess pulse pressure applied to the fluid in the dispersed phase reservoir. A LabVIEW program is used to maintain a constant pressure of the dispersed phase to maintain the fluid meniscus at the intersection of the T-junction and to apply a pulse of high pressure for a short duration to generate droplets when necessary using manual control with a Boolean switch. The actual pulse durations and pressures are tuned depending on the oil-aqueous solutions being studied.

Flow through the cross-slot microfluidic device is visualized using a 10% v/v emulsion of water in light mineral oil. Open-source PIV software (PIVlab) is used to extract flow-field information [73]. The flow field in the cross-slot contains a stagnation point near the center of the cross-slot, as seen in **Figure S2**. Particles and droplets are trapped at the stagnation point in a Stokes trap using the Model Predictive Control (MPC) strategy implemented previously by Shenoy et al. [51]. Briefly, droplets that enter the region of interest, defined here as the cross-slot region, are detected using binary image processing and particle tracking in LabVIEW. To steer a droplet from its initial position to the final desired position – here, the center of the cross-slot – an optimum trajectory is determined, such that the flow rates are reasonable and incremented in small steps, and the particle reaches the set target position via the shortest possible trajectory. To achieve this, an optimization problem is solved by discretizing a finite time period, referred to by Shenoy et al. [51] as a finite time horizon, into intervals with a frequency equal to the sampling frequency. In this case, the sampling frequency of the image is the camera frame rate, which is set to 30 fps for the Basler ace

acA1300-60gm camera (max. frame rate is 60 fps). This MPC algorithm solves for the flow rates required to steer the droplet along the shortest trajectory and is iterated after every time interval. It is implemented using the Automatic Control and Dynamic Optimization (ACADO) toolkit for solving non-linear MPC problems [74].

Consider that the microfluidic cross-slot has $i$ channels, where $i = 4$ for the 4-channel cross-slot. The flow rates $q_i$ in each channel of the cross-slot are calculated using the MPC algorithm. These flow rates are converted to pressures $p_i$ using the relation $p_i = p_0 + r_i q_i$, where $p_0$ is a base pressure (set to 3 psi (gauge) in most cases), $r_i$ denotes the hydrodynamic resistance in the channel, which is calculated using the formula $r_i = \frac{12\eta_c L_i}{1 - 0.63\left(\frac{h}{w}\right)} \frac{1}{h^3 w_i}$ [48], where $\eta_c$ is the continuous phase viscosity and $L_i, h$ and $w_i$ are the lengths, height and widths of the channels respectively. The Stokes trap for particles implemented by Shenoy et al. is a symmetric device with all the channels and tubing being of equal length. In the case of our modified Stokes, trap, the resistance of the microfluidic channel containing the droplet formation region is almost five times as high as that of the other three channels forming the cross-slot. Moreover, the shear rate required in the T-junction to form droplets is high, requiring high pressures to be applied to the channels of the T-junction. This gives rise to high velocity of droplets in the trap region with a net flow of droplets into the channel P1 (**Figure 2c**). Therefore, rapid response of the trapping algorithm is required to capture droplets. Shenoy et al. [51,75] define the optimization problem for the flow rates in terms of two controller weights, denoted as $\beta$, which relates to flow rates in the channels and $\gamma$, which is a 'compensation' for deviation from the set position. In general, lower values of $\beta$ ($10^{-5}$ to $10^{-4}$) and higher values of $\gamma$ (>1000) are found to be appropriate for trapping droplets in this experiment. These controller weights can be tuned for fluids of different viscosities.

*2.2 Materials*

In this work, we use two different continuous phases (light and heavy mineral oils, Sigma Aldrich) to vary the continuous phase viscosity. Aqueous glycerol solutions at 0%, 20%, 50% and 70% by volume prepared using HPLC grade water (Fisher Scientific) and glycerol (Fisher Scientific) are used as the dispersed (droplet) phase for droplet deformation experiments. A nonionic surfactant, SPAN 80 (Sorbitane monooleate, Croda Inc.) at a fixed concentration of 0.05% v/v is added to the continuous phase, i.e. oil, to achieve uniform droplet formation. The equilibrium interfacial tensions of the oil-aqueous systems are measured using pendant drop tensiometry with a Drop Shape Analyzer (Krüss GmbH) at room temperature using standard protocol [76]. Note that the equilibrium interfacial tensions of all the systems studied here lie in the range of 5-10 mN/m as indicated in **Figure 3**.. These systems are chosen so as to have minimal changes in interfacial tension but large changes in viscosities. The viscosities of the aqueous glycerol solutions and the mineral oils are measured using an AR-G2 rotational rheometer (TA instruments) with a concentric cylinder geometry over a shear rate range of 1 – 1000 s$^{-1}$, and the fluids are found to be Newtonian over this shear rate range. The experiments are performed with light and heavy mineral oils in the continuous phase with dynamic viscosities ($\eta_c$) of 26.8 mPa-s and 132.2 mPa-s respectively. The dispersed phases used include water and glycerol-water solutions with viscosities ($\eta_d$) in the range 1.73 – 23.13 mPa-s. The viscosity ratios ($\lambda = \frac{\eta_d}{\eta_c}$) range from 0.007 to 0.87, spanning almost two orders of magnitude.

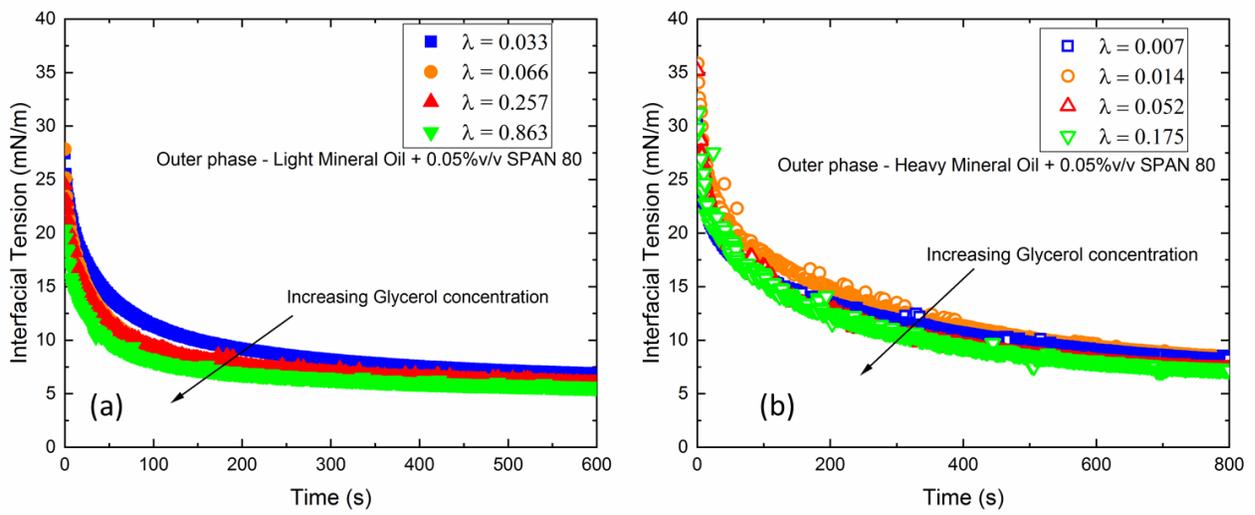

**Figure 3:** (a) Interfacial tension measurements using pendant drop tensiometry for light mineral oil with 0.05% v/v SPAN 80 in the continuous phase with 0-70% aqueous glycerol in the dispersed phase. (b) Interfacial tension measurements using pendant drop tensiometry for heavy mineral oil with 0.05% v/v SPAN 80 in the continuous phase with 0-70% aqueous glycerol in the dispersed phase.

## 3. Results and Discussion

*3.1 Droplet deformation in a microfluidic hydrodynamic trap*

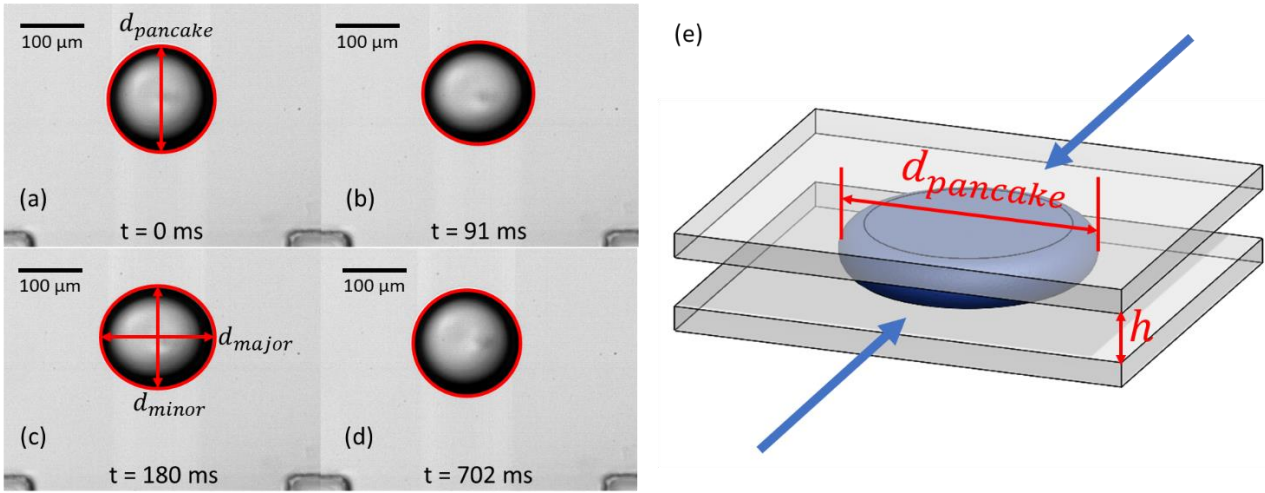

**Figure 4:** (a) – (d) Droplet shape perturbation after application followed by cessation, of a pressure pulse at various instants of time, showing deformation and relaxation of droplet shape. The original un-deformed projected droplet diameter $d_{pancake}$ and the lengths of the major and minor axes of the deformed droplet ($d_{major}$ and $d_{minor}$) are indicated in panels (a) and (c), respectively. (e) Sketch of the pancake-shaped droplet confined by the walls of the microfluidic device, showing the direction of the applied pressure pulse (blue arrows) and the channel height $h$ compared with the projected drop diameter $d_{pancake}$.

Droplets trapped at the center of the cross-slot in the microfluidic trap are deformed by applying an excess pressure pulse of 10 psi to channels P2 and P4 (see **Figure 2a-c**). A pulse pressure of 10 psi is chosen so as to induce measurable deformation of a trapped droplet while also keeping it trapped at the center of the cross-slot. Projected droplet diameters $d_{pancake}$ (measured in top view) are in the range 90 – 250 µm, with the device height $h$ being ~100 µm on average (due to PDMS channel swelling by mineral oils). Device height after swelling is measured using the calibrated Z-stage of the inverted microscope. Prior to the application of excess pressure, the droplet is in its equilibrium configuration and has a 'pancake' shape with a circular base. The drop is not pinned against or stuck to the channel floor and ceiling, and a lubrication layer of the continuous phase separates the drop from the walls. After application of the pressure pulse, the droplet assumes a pancake shape with an elliptical base as shown in **Figures 4b and c**. The pulse is applied for a period of 50 milliseconds, after which the excess pressure is reduced to zero psi, while the droplet is still trapped. This allows the droplet to relax back to its equilibrium configuration. The droplet is imaged in 2-D (i.e. in top view) for approximately 1 second, such that it appears circular prior to deformation and elliptical after the pressure pulse is applied. High speed videos of droplet shape relaxation are captured at a frame rate of 12,500 fps using the Photron Mini UX 100 camera. These videos are then analyzed in MATLAB using a custom image analysis code [6,14] which detects droplet diameter, major and minor axes lengths. The deformed shape of the droplet is described by a deformation parameter '$D$', coined by G.I. Taylor in his pioneering studies on droplet deformation in emulsions [17,18]. Here,

$$D = \frac{d_{major} - d_{minor}}{d_{major} + d_{minor}}, \tag{1}$$

where $d_{major}$ and $d_{minor}$ are the lengths of the major and minor axes of the ellipse respectively.

The image analysis yields the droplet deformation parameter $D$ as a function of time, which is obtained from the frame rate (here, 12,500 fps). **Figure 5** shows $D$ as a function of time for the system consisting of light mineral oil in the continuous phase with 50% glycerol-water solution in the dispersed phase, for droplets of various diameters. Initially, $D$ sees a sharp increase as the droplet is deformed, and the droplet gradually relaxes back to its equilibrium shape when the pressure pulse is stopped. An exponential decay is fit to the drop shape relaxation curve starting at the maximum observed deformation, and is given by

$$D = D_0 + A e^{-\frac{t}{\tau}}, \tag{2}$$

where $D_0$ is the equilibrium deformation, $A$ is the deformation at time $t = 0$ (chosen to be the point of maximum deformation) and $\tau$ is the relaxation time for droplet shape measured from experiments.

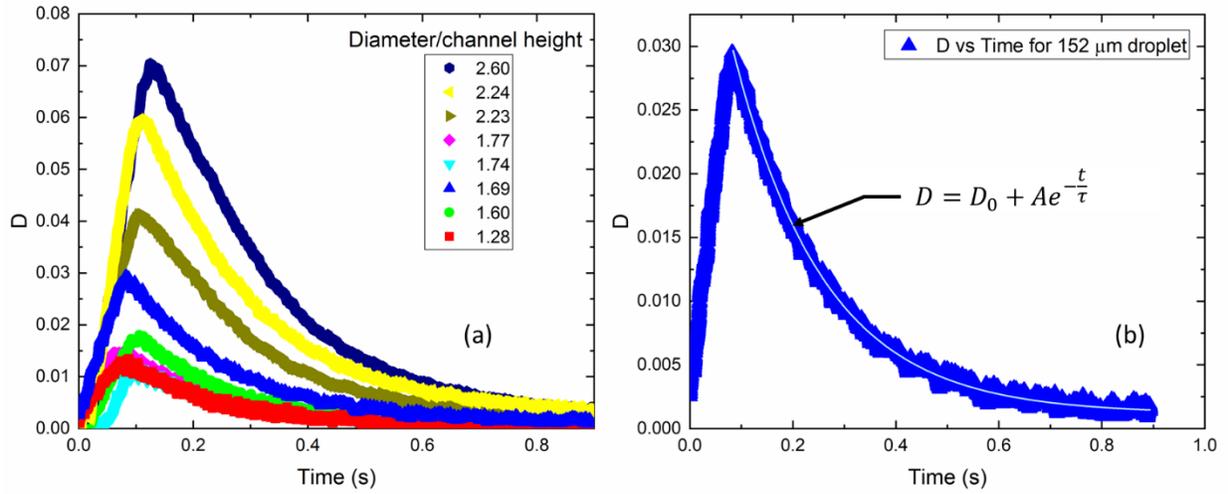

**Figure 5:** (a) Droplet deformation '*D*' as a function of time for the system containing 50% glycerol-water solution in the dispersed phase with light mineral oil in the continuous phase. Different symbols indicate droplets with different initial diameters. Droplets attain highly deformed shapes, followed by relaxation back to equilibrium. (b) Exponential decay fit to droplet shape relaxation for a 152 μm diameter droplet of 50% glycerol-water solution in light mineral oil.

When a droplet is deformed by the pressure pulse, the viscous stress in the continuous phase acts to deform the droplet, while interfacial tension between the two liquid phases acts to restore droplet shape. Similarly, as the droplet shape relaxes back to equilibrium, the interfacial tension and viscous stress in the continuous phase act as competing forces. Therefore, it is obvious that the continuous phase viscosity must influence the relaxation time. On the other hand, when the continuous phase is much more viscous than the dispersed phase, the dispersed phase viscosity is not considered as a major parameter influencing the droplet's shape relaxation. Here, we test this hypothesis by changing the dispersed phase viscosity relative to the continuous phase. **Figure 6(a) and 6(b)** show the relaxation time measured for a range of droplet sizes, with 0%, 20%, 50% and 70% glycerol-water solutions in the dispersed phase and light and heavy mineral oils in the continuous phase respectively. Across the entire range of viscosity ratios, it is found that the relaxation time increases with an increase in droplet diameter. This is because under high confinement, it is not just the viscous stress, but the pressure in the continuous phase, which depends on the square of the drop radius, that affects drop dynamics [62]. With light mineral oil in the continuous phase, it is also observed that the relaxation time exhibits no dependence on the dispersed phase viscosity (or the viscosity ratio between the two phases) in the low viscosity ratio limit as shown in **Figure 6(a)**. However, for the 70% aqueous glycerol solution in light mineral oil, the viscosity ratio increases substantially compared to all the other cases ($\lambda = 0.87$) as the viscosity of the droplet phase nearly equals that of the continuous phase. In this case, we observe that the droplet shape relaxation time is notably lower. This decrease in relaxation time indicates that for a highly viscous droplet, the energy dissipation inside the droplet (in the boundary layer on the droplet side) is extremely rapid, leading to rapid return to the equilibrium shape. Similarly, as shown in **Figure 6(b)**, with heavy mineral oil in the continuous phase, the relaxation time is smaller for the highest viscosity ratio of 0.175, although the difference is not as evident as is the case with light mineral oil in the continuous phase. **Figures 6(c) and (d)** show the maximum deformation as a function of droplet diameter over the entire range of viscosity ratios. Here, it is observed that the maximum deformation increases with increase in droplet size over the entire range of dispersed phase viscosities. However, the maximum deformation does not show any dependence on viscosity ratio.

The role of the dispersed phase in the dissipation of shape deformation is discussed in literature dating back to work by Lamb [16] and Chandrasekhar [19], which examines shape oscillation and damping of a viscous liquid sphere in a fluid of negligible viscosity. These derivations show that oscillations are suppressed for a viscosity exceeding a 'critical viscosity' defined by Chandrasekhar [19], and droplet shape relaxes exponentially to equilibrium. For a drop suspended in a fluid of negligible viscosity (e.g. in air), the relaxation time varies inversely as the droplet's viscosity. This theoretical treatment has also been applied experimentally by Bzdek et al [32] for the measurement of viscosity of optically trapped aerosol droplets. Drawing an analogy to the aforementioned theories for viscous droplets in air, it is reasonable to expect that when the continuous phase viscosity is kept constant and the dispersed phase viscosity increased, the relaxation time for droplet shape would decrease with an increase in dispersed phase viscosity, as seen in the case of 70% aqueous glycerol solution in light mineral oil.

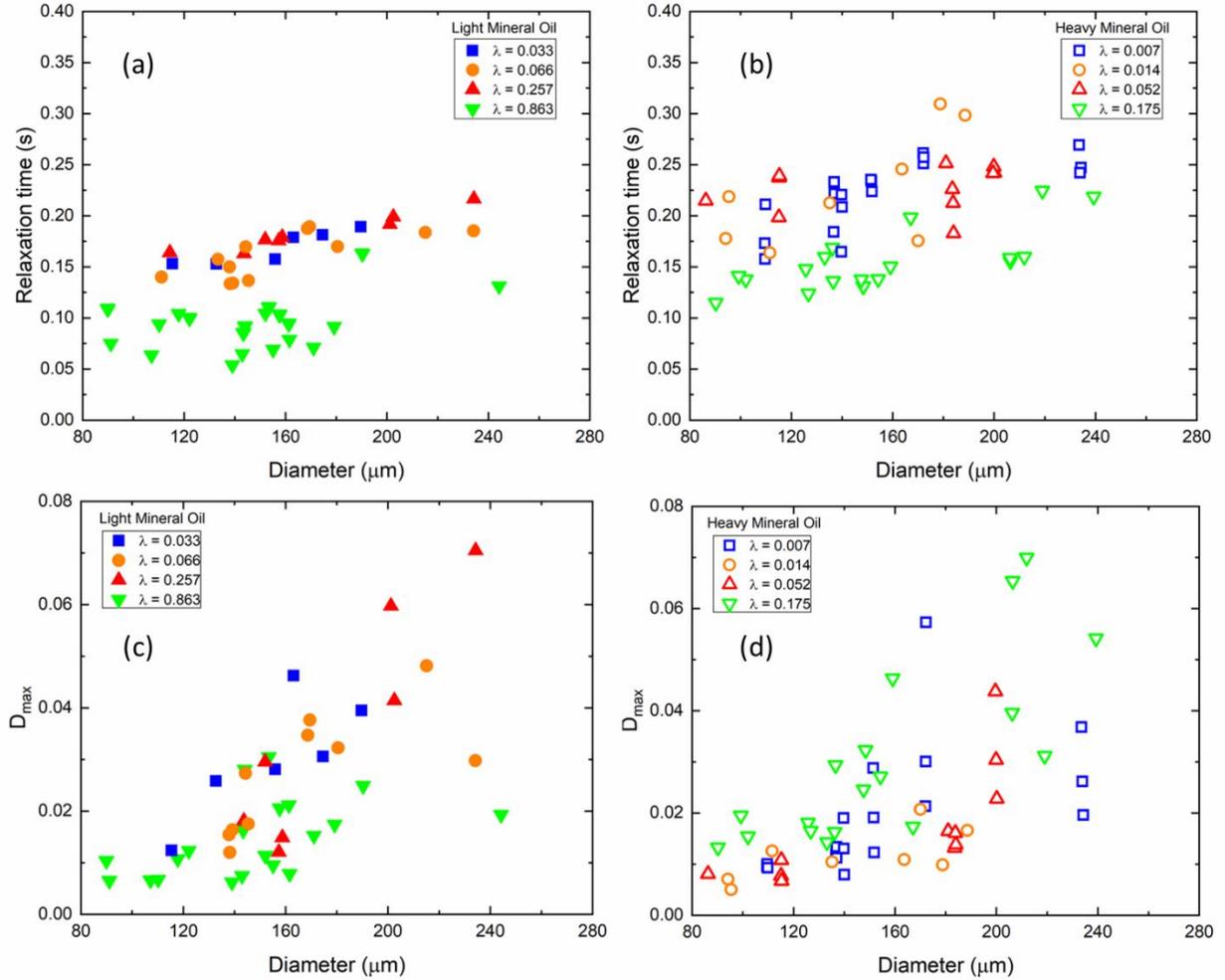

**Figure 6:** (a), (b) Experimentally measured droplet shape relaxation time as a function of droplet diameter for (a) light mineral oil and (b) heavy mineral oil in the continuous phase with 0%, 20%, 50% and 70% aqueous glycerol in the dispersed phase (c), (d) Maximum experimentally observed deformation as a function of droplet size for (c) light mineral oil and (d) heavy mineral oil in the continuous phase with 0%, 20%, 50% and 70% aqueous glycerol in the dispersed phase.

*3.2 Theory for droplet shape relaxation in unconfined and confined flows*

The characteristic timescale for a free droplet's relaxation to equilibrium was derived by Taylor and others [18,30,41,77] for an initially spherical droplet in an unconfined flow field in the limit of very small to moderate viscosity ratios. This timescale, called the viscous dissipation time, is a function of the droplet radius ($r = \frac{a_0}{2}$), viscosity ratio ($\lambda$), interfacial tension ($\gamma$), and either the continuous ($\eta_c$) or dispersed ($\eta_d$) phase viscosity. Additionally, $a_0$ is the diameter of a spherical droplet having the same volume as a pancake-shaped confined droplet with a projected diameter $d_{pancake}$, and is calculated using the formula $a_0 = \sqrt[3]{\frac{h^3}{6} + \frac{3h}{2}(d_{pancake} - h)\left(\frac{\pi h}{2} + d_{pancake} - h\right)}$ [78,79], where $h$ is the height of the microfluidic channel. $d_{pancake}$ is obtained from the measured projected diameter of the confined droplet from the experiments, and the calculated diameter $a_0$ is the diameter of an unconfined spherical droplet having the same volume as the confined pancake-shaped droplet when $d_{pancake} > h$ and is taken to be equal to $d_{pancake}$ when $d_{pancake} < h$. Here, we define the viscous dissipation timescale in terms of the continuous phase viscosity to illustrate order-of-magnitude trends compared with the experimentally measured relaxation time. However, it is important to be aware of the viscosity ratio range ($\lambda \ll 1$) in which this theoretical viscous dissipation time ($\tau_{visc-c}$) is valid [30]. It is given by

$$\tau_{visc-c} = \frac{\alpha \eta_c r}{\gamma}, \tag{3}$$

where the parameter $\alpha$ is a function of the viscosity ratio $\lambda$, and is given by

$$\alpha = \frac{(2\lambda + 3)(19\lambda + 16)}{40(\lambda + 1)} \tag{4}$$

**Figure 7(a)** shows the characteristic relaxation time for droplets over a range of drop sizes (same volume as the range investigated in this work) calculated using Equation (3). It can be seen that in the unconfined case, i.e. for a droplet relaxing to equilibrium in a quiescent, infinite fluid reservoir, the droplet relaxation time is driven by two key parameters, namely the continuous phase viscosity and the interfacial tension. Since the viscosity of light mineral oil is lower, the relaxation time is smaller than the heavy mineral oil cases. Moreover, it is interesting to note that the viscosity ratio does not strongly affect the relaxation time when the viscosity ratio is small ($\lambda \ll 1$) as seen in **Figure 7(b),** whereas the curve begins to deviate and increase substantially as the viscosity ratio approaches 1.

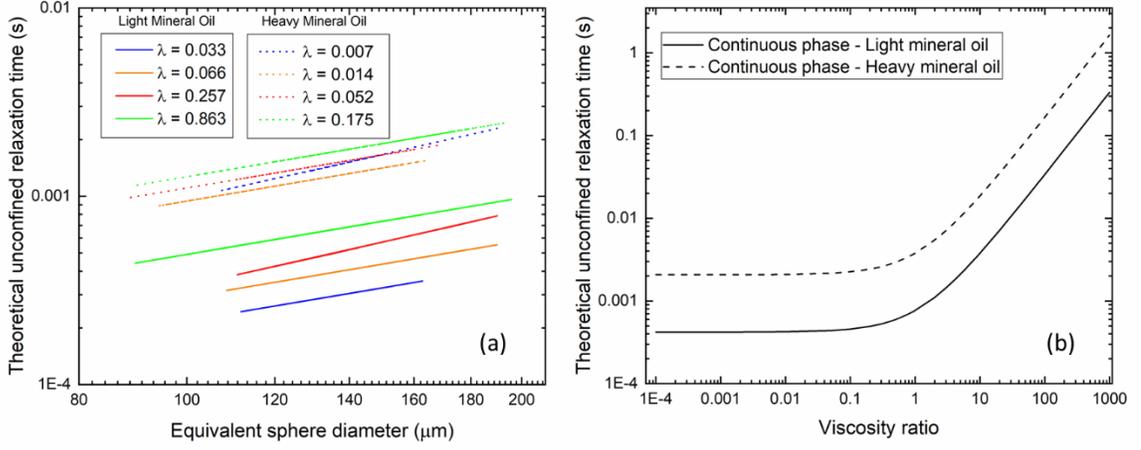

**Figure 7:** (a) Theoretical relaxation time for unconfined droplets as a function of droplet diameter over the range of droplet diameters and viscosity ratios in this study (b) Theoretical relaxation time for unconfined droplets as a function of viscosity ratio for a droplet of equivalent sphere diameter 157 μm with an interfacial tension of 6 mN/m.

In the highly confined flows relevant to the experiments in this study, the relaxation of droplet shape is strongly affected by the presence of the walls of the microfluidic device. The droplet is aqueous and does not wet the hydrophobic device walls. Therefore, there exists a thin film of oil between the droplet and the top and bottom walls of the device. A numerical scheme for the shape evolution of a pancake-shaped droplet in microfluidic channels was developed by Gallaire and coworkers [63–65]. For shallow channels, the velocity profile in the Z-direction (in our experiments, this is perpendicular to the focal plane) is assumed to be parabolic, and the droplet's deformation can be considered to be primarily occurring in the X-Y plane (in our experiments, this is the focal plane of the objective). A model for analyzing flow through porous media, known as Darcy's law, is modified to include a correction for the depth-averaged Laplacian, yielding the so-called Brinkman equation [63,64,80–82], given by

$$\lambda \left( \Delta \boldsymbol{u} - k^2 \boldsymbol{u} \right) = \nabla p \qquad (5)$$

$$\nabla \cdot \boldsymbol{u} = 0 . \qquad (6)$$

Here, $k = \sqrt{12}\frac{r_{pancake}}{h}$ where $h$ is the height of the microfluidic channel, $r_{pancake} = \frac{d_{pancake}}{2}$, $p$ is the pressure field and $\boldsymbol{u}(x,y)$ is the two-dimensional velocity vector. Note that pressure and velocity are non-dimensionalized by $\tilde{p} = \frac{\gamma}{r_{pancake}}$ and the capillary velocity $\tilde{v} = \frac{\gamma}{\eta_d + \eta_c}$ respectively.

The Brinkman equations were solved numerically by Gallaire and coworkers [63–65] using the boundary element method. This numerical scheme was successfully applied for droplets entering a microfluidic expansion geometry, and the results compare well with experimental work by Brosseau et al. [64] Similarly, results from this method were compared with experiments for droplets flowing through a cross-slot by Ulloa et al. [62] and good agreement was obtained. Brun et al. [63] also investigated the relaxation of a stationary droplet inside a microfluidic channel, and found that a rescaling of the time $t$ as $t' = t\frac{h^2}{r_{pancake}^2}$ leads to collapse of the droplet shape relaxation plots.

*3.3 Comparison of experimental and theoretical relaxation times*

**Figure 8(b)** shows the deformation $D$ plotted as a function of the rescaled time $t' = t\frac{h^2}{r_{pancake}^2}$ for 50% glycerol-water droplets in light mineral oil, according to the results of the boundary element method simulation by Brun et al. [63]. All the deformation curves collapse after rescaling, indicating that the decay time is indeed proportional to the square of the radius, in agreement with the boundary element simulations. Here, the channel height is assumed to be 100 μm to account for swelling of PDMS by mineral oils.

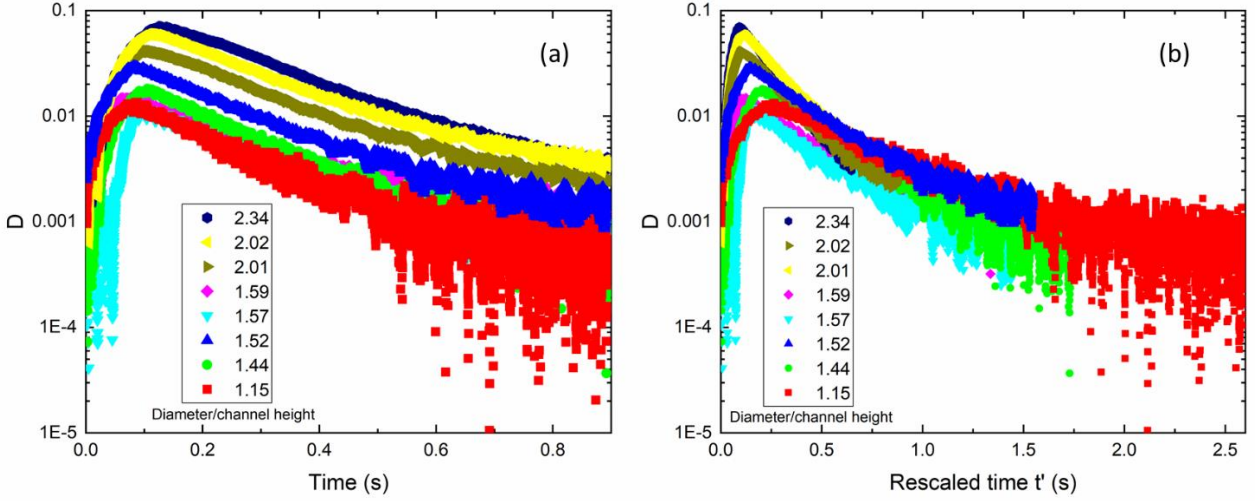

**Figure 8:** (a) Droplet deformation versus time for the light mineral oil – 50% aqueous glycerol system at various droplet sizes (b) Droplet deformation $D$ as a function of rescaled time $t' = t \frac{h^2}{r_{pancake}^2}$, for the same system. Symbols and colors indicate droplet size. Collapse of deformation curves for various droplet sizes is obtained with this rescaling of the time axis.

Further, the dependence of the droplet's relaxation time on the dispersed phase viscosity obtained from our microfluidic experiments is compared with the results obtained by numerical simulation. Brun et al. [63] found that the droplet's relaxation time varies inversely as the square of the projected radius according to the relation

$$\tau_n \sim -\frac{n(n^2-1)\pi}{4k^2} \sim -\frac{n(n^2-1)\pi h^2}{48\, r_{pancake}^2}\ . \tag{7}$$

Here, $n$ is a constant denoting the Fourier mode of shape deformation. Therefore, $\tau_n \propto \frac{h^2}{r_{pancake}^2}$. As per the above equation, it is evident that the experimentally measured relaxation times for a pancake-shaped droplet should scale as $\frac{h^2}{r_{pancake}^2}$. Moreover, Ulloa et al [62] found that for highly confined droplets, the pressure in the outer fluid, which scales as $\frac{\eta_c r_{pancake}^2}{h^2}$ at a constant strain rate, strongly affects drop deformation. **Figure 9(a) and (b)** show the experimentally measured relaxation times plotted as a function of $\frac{h^2}{r_{pancake}^2}$ for all glycerol-water ratios in the dispersed phase and light and heavy mineral oils in the continuous phase respectively. It is evident that the experimentally measured relaxation times do indeed vary inversely as $\frac{h^2}{r_{pancake}^2}$. **Figure 9(a)** reveals that the experimentally measured relaxation time exhibits a weak dependence on viscosity ratio when $\lambda \ll 1$. However, as discussed before, when $\lambda \to 1$, i.e. in the case of 70% aqueous glycerol in light mineral oil, the experimentally measured relaxation time is markedly different. **Figure 9(c)** shows the relaxation time as a function of viscosity ratio for a pancake-shaped droplet of diameter ~ 180 μm. A decreasing trend in relaxation time is observed with viscosity ratio, indicating that indeed the relaxation time depends weakly on viscosity ratio when the interfacial tension is kept relatively constant.

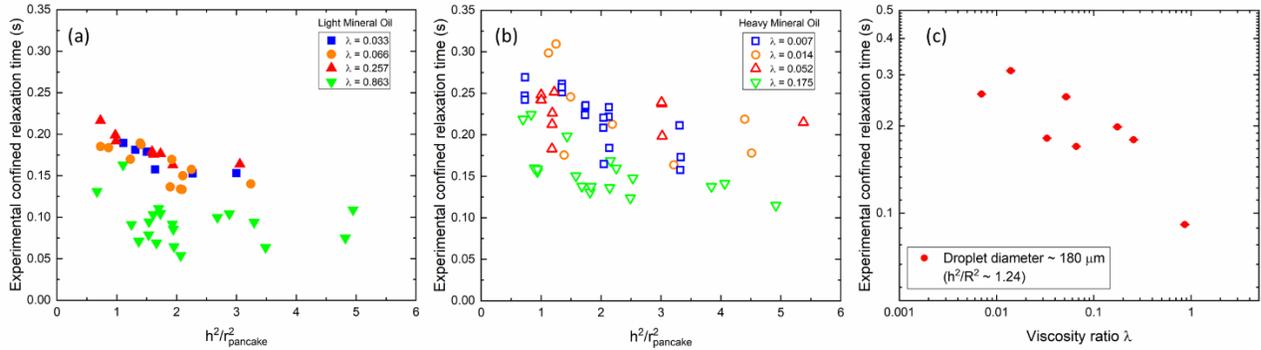

**Figure 9:** (a) Measured relaxation times as a function of $\frac{h^2}{r_{pancake}^2}$ for all systems studied here. Open symbols indicate heavy mineral oil (HMO) in the continuous phase, and closed symbols indicate light mineral oil (LMO) in the continuous phase. Shapes/colors of symbols indicate different viscosity ratios. (b) Experimental relaxation times as a function of viscosity ratio for droplet diameter ~ 180 μm.

A comparison of **Figures 6(a) and 7(a)** sheds light on an important difference between the experimentally measured droplet shape relaxation times and the theoretical relaxation time for free, unconfined droplets. The relaxation time for unconfined droplets ($\tau_{visc-c}$) of the same diameters and with the same continuous and dispersed phases and interfacial tensions is approximately 2 orders of magnitude or 100 times smaller than the experimentally measured relaxation times for confined, pancake-shaped droplets. This observation indicates that the presence of the confining walls (or the top and bottom walls of the microfluidic channel) has a significant impact on the drop shape relaxation dynamics. This is qualitatively in agreement with previous work investigating droplet shape relaxation in confined flows. For example, Vananroye et al. [58] studied the relaxation of polydimethylsiloxane droplets in polyisobutylene upon cessation of shear flow and found that highly confined droplets relax much slower than unconfined or moderately confined droplets. Moreover, for extensional flow around a confined, moving droplet, Brosseau [77] found that the experimental relaxation time lags behind the theoretical relaxation time by an order of magnitude. Tiwari et al [83] also found that the relaxation time of droplets following pinch-off of a liquid bridge in a Hele-Shaw cell increases with increase in confinement.

For confined droplets, the drag force exerted by the confining walls is thought to have an effect on the droplet relaxation, and in general, causes the relaxation to lag behind that for an unconfined droplet. As mentioned earlier, it is well-known that for micro-confined droplets that do not wet the device walls, a thin Bretherton film exists between the droplet and the wall, and its thickness has been estimated to be within 0.1-3 µm for pancake-shaped droplets in microchannels [84–87]. The existence of the continuous phase fluid within this thin lubrication film dictates that the continuous phase viscosity must have an impact on the droplet shape relaxation time, which is indeed the case as shown in **Figure 9(a) and (b)**, where the relaxation time with heavy mineral oil in the continuous phase is generally higher than that for light mineral oil in the continuous phase. A pancake shaped droplet in a microchannel can be approximated as a combination of a cylinder and two endcaps. The radius of the endcaps can be assumed to be the difference between the half-width of the channel ($h/2$) and the thickness of the Bretherton film ($h_\infty$). Therefore, the radius of the area forming the lubrication film with the channel wall can be estimated as $r_c = r_{pancake} - (\frac{h}{2} - h_\infty)$. The drag force on the droplet should therefore scale as the contact area of the droplet i.e. as $r_c^2 \sim r_{pancake}^2$. This additional forcing due to wall drag would slow down the relaxation of the droplet and would increase with droplet size as well as continuous phase viscosity. In other words, the droplet would relax slowly if the continuous phase is more viscous (e.g. heavy mineral oil) or if the droplet has a larger projected radius. A full hydrodynamic analysis of the drag force on confined droplets as a function of inner and outer fluid viscosity is outside the scope of this work, but similar analysis for bubbles in microconfined flows can be found in previous work by Rabaud et al. [88].

*3.4 Outlook for droplet shape relaxation during coalescence*

Droplet shape relaxation studies find application in several interesting problems in multiphase flows. One such problem is droplet coalescence in micro-confined environments. Here, a single droplet is trapped at the center of the hydrodynamic trap, and a series of incoming droplets are allowed to coalesce with the trapped droplet. **Figures 10(a)-(d)** show the coalescence of water droplets in light mineral oil with 0.01% v/v SPAN 80 in the continuous phase. Immediately following coalescence, the coalesced droplet attains a highly distorted geometry, which ultimately relaxes back to a spherical shape as shown in **Figure 10(d)**. These highly distorted shapes and their relaxation back to a spherical shape can be strongly influenced by the fluid properties, surfactant gradients after coalescence, as well as the confinement of the droplet in the microfluidic device. Further, using the hydrodynamic trap, it is possible to measure the film drainage time for droplet coalescence. Droplet coalescence in linear flows occurs in three stages, viz. (1) approach of the droplets from a starting separation distance (2) contact between droplets and drainage of the thin film between them, during which droplets may rotate about a common axis (3) rupture of the thin film, leading to coalescence and formation of highly distorted droplet shapes which eventually relax to a spherical shape [89]. It is well-documented in the literature that the film drainage time for droplet coalescence is a function of the droplet velocity, surfactant concentration, droplet size and fluid properties [89–91]. Future work using the hydrodynamic trap for droplet coalescence will involve a detailed investigation of these parameters and their effect on film drainage time and droplet shape relaxation during coalescence.

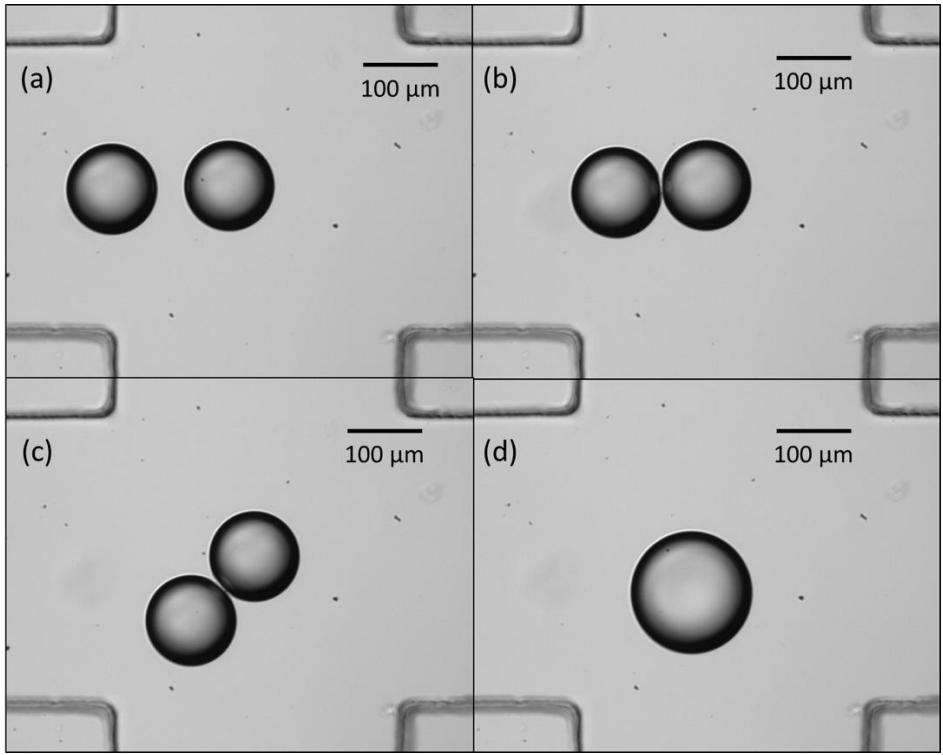

**Figure 10**: (a) – (d) Coalescence of an incoming droplet with a droplet trapped at the center of the cross slot showing the various stages of coalescence. The system shown here consists of confined water droplets in light mineral with 0.01% v/v SPAN 80.

## 4. Conclusions and Outlook

In this work, a 'Stokes trap' previously developed by Shenoy et al. [51] is modified for two-phase liquid-liquid systems to include on-chip droplet formation using a drop-on-demand technique. Droplets composed of aqueous glycerol solutions of varying viscosities are formed in light and heavy mineral oils, trapped at the center of the cross-slot and perturbed to induce a sudden shape deformation. After the pressure pulse is removed, the droplet shape relaxation time is measured from the drop shape decay curves. It is found that the droplet shape relaxation time depends weakly on the viscosity ratio $\lambda$. Moreover, it is found that the continuous phase viscosity affects the droplet relaxation time, resulting in higher relaxation times for more viscous oil in the continuous phase. This is because the pressure in the continuous phase, which scales as continuous phase viscosity and the square of the ratio of drop radius to channel height at a given strain rate, acts to deform the droplet, while interfacial tension acts to restore droplet shape. A comparison of experimentally measured relaxation times for confined droplets with theoretical relaxation times for unconfined droplets reveals that the experimentally measured relaxation times for confined droplets of the same volume are orders of magnitude larger than unconfined case. This is because in confined environments, there is an additional forcing due to drag from the top and bottom channel walls, which varies as the square of the projected radius of the droplet. The confinement of the droplet is, therefore, found to be an extremely significant contribution to the droplet shape relaxation dynamics over the range of viscosity ratios and drop sizes studied here.

The results of this work have implications for the behavior of confined droplets in various emulsion processing applications including transients such as flow start-up and sudden cessation, leading to relaxation of droplet shape. The results obtained indicate that while the viscosity ratios have a small impact on drop shape relaxation, the effect of confinement is much more dominant compared to unconfined droplets. Previous work has also shown that the presence of surfactants can lead to Marangoni effects along the droplet interface during deformation and may impart a dilatational elasticity to the interface [92–94]. A rigid droplet interface may also exhibit different relaxation behavior. It is unclear whether these gradients can strongly influence the drop dynamics, particularly under high degrees of confinement. Future work with micro-scale emulsion droplets will investigate the effect of varying surfactant concentration on the interfacial dilatational modulus, and any potential impact on droplet deformation and coalescence.

Finally, the microfluidic hydrodynamic trap for droplets developed in this study can be applied towards studying the dynamics of droplet coalescence in a precisely controlled flow environment, with a range of droplet sizes, surfactant concentrations and droplet speeds. The droplet shape relaxation dynamics studied in this work can be extended to study the post-coalescence relaxation of droplets. Film drainage times for coalescence can also be measured as a function of various influencing parameters. The droplet-based microfluidic hydrodynamic trap is therefore a versatile lab-on-a-chip platform for studying fundamental single droplet dynamics and droplet-droplet interactions at length and time-scales relevant to commercial applications of liquid-liquid systems.

**Acknowledgements**

This work was funded by Donaldson Company (Bloomington, MN) and carried out at the University of Minnesota. We would like to thank Prof. Charles Schroeder, Dr. Anish Shenoy and Mr. Dinesh Kumar from the University of Illinois at Urbana-Champaign (Chemical and Biomolecular Engineering, Mechanical Engineering), for help with setting up the Stokes trap and providing the code for MPC-based feedback control. We would also like to thank Dr. Benjamin Micklavzina (University of Minnesota) and Prof. Andrew Metcalf (Clemson University) for helpful discussions. Rheological measurements were conducted at the Polymer Characterization Facility and pendant drop tensiometry measurements were conducted at the Coating Process Fundamentals Lab, both in the Chemical Engineering and Materials Science department at the University of Minnesota. Portions of this work were conducted in the Minnesota Nano Center, which is supported by the National Science Foundation through the National Nano Coordinated Infrastructure Network (NNCI) under Award Number ECCS-154220.

**Supporting Information for**

**A 4-channel microfluidic hydrodynamic trap for droplet deformation and coalescence in extensional flows**

*Shweta Narayan[1], Davis B. Moravec[2], Andrew J. Dallas[2] and Cari S. Dutcher[1]\**

*[1]Department of Mechanical Engineering, University of Minnesota – Twin Cities, MN 55414*

*[2]Donaldson Company Inc., Bloomington, MN 55431*

*\*Corresponding author*

1) **Profilometry measurements**

A KLA-Tencor P-16 Profilometer is used to measure the height of the microfluidic channels after the features are created by soft lithography on the silicon wafer, and prior to pouring PDMS on the wafer.

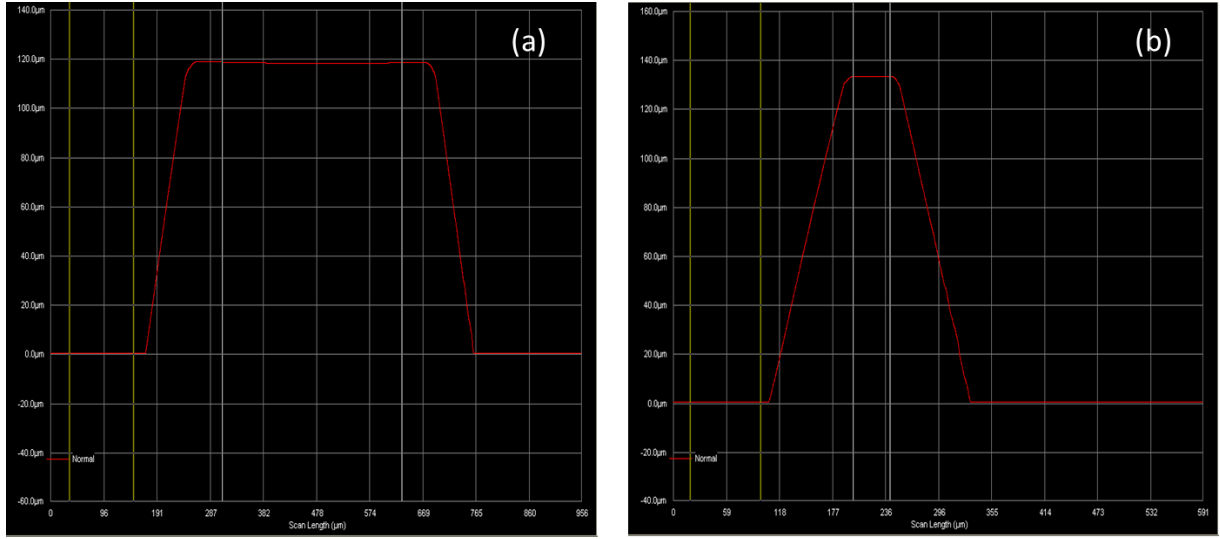

**Figure S1**: Profilometry measurements showing the height of the device prior to pouring PDMS at (a) the Stokes trap region, where the height is 118.2 μm and (b) the T-junction for droplet formation, where the height is 133.2 μm. Note that non-uniformities in height may occur due to uncertainties during the spin-coating of photoresist onto the silicon wafer. Furthermore, injection of mineral oils causes the PDMS device to swell, resulting in the effective height being lower than the original height.

2) **Particle Image Velocimetry (PIV) analysis**

Micro-PIV measurements are conducted by injecting a 10% v/v water-in-oil emulsion into the hydrodynamic trap device. The continuous phase is light mineral oil with surfactant SPAN 80 added at 0.05% v/v for stabilizing the emulsion. The device is mounted on an Olympus IX83 inverted microscope and imaged using bright-field microscopy. Images are taken in the center plane of the device along the height (horizontal plane) using a high-speed camera (Photron Mini UX100) at a frame rate of 12,500 fps. The images are analyzed using the PIVlab tool for MATLAB [1,2].

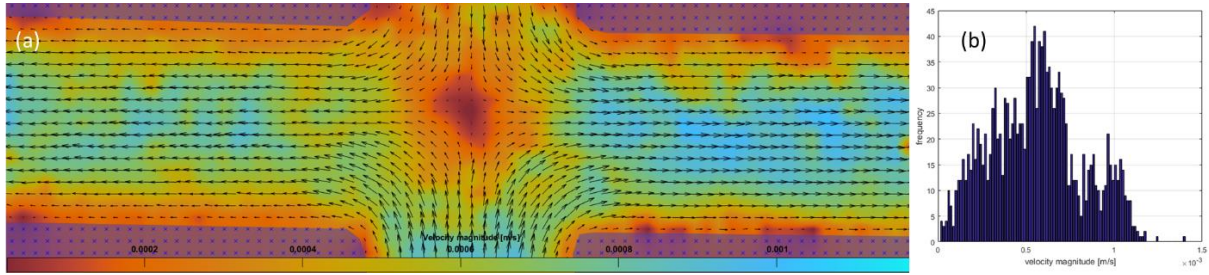

**Figure S2:** (a) PIV analysis of the flow field in a hydrodynamic trap showing velocities and the location of the stagnation point at the center of the cross-slot. (b) Velocity distributions in the microfluidic trap region.

3) **Viscosity measurements**

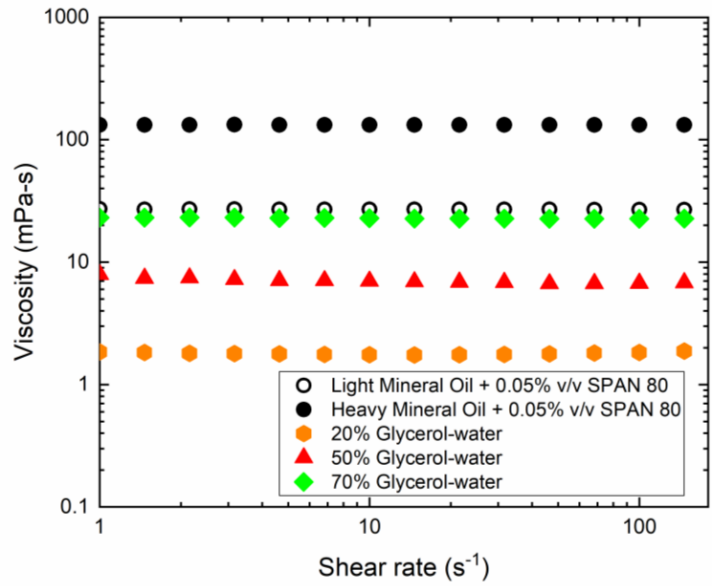

**Figure S3:** Viscosity as a function of shear rate for the mineral oils and aqueous glycerol solutions in this study, measured using an AR-G2 rotational rheometer (TA Instruments) with a concentric cylinder geometry.